\documentclass[%
 reprint,
superscriptaddress,
preprintnumbers,
 amsmath,amssymb,
 aps,
prb,
twocolumn,
floatfix,
]{revtex4-1}
\usepackage{amsfonts}
\usepackage{amsmath}
\usepackage{amssymb}
\usepackage{hyperref}
\usepackage{graphicx}% Include figure files
\usepackage[export]{adjustbox}
\usepackage{dcolumn}% Align table columns on decimal point
\usepackage{bm}% bold math
\usepackage{tikz}
\usepackage{braket}
\newcommand{\hamil}{\mathcal{H}}

\begin{document}

\title{Classical spin models and basic magnetic interactions on 1/1-approximant crystals
}% Force line breaks with \\
%\thanks{This work is funded by the Knut and Alice Wallenberg Foundation}%

\author{Daniel Qvarngård}
 \email{danielqv@kth.se}
\affiliation{Department of Physics, Royal Institute of Technology, SE-106 91 Stockholm, Sweden}
\author{P. Henelius} 
\affiliation{Department of Physics, Royal Institute of Technology, SE-106 91 Stockholm, Sweden}
\affiliation{Faculty of Science and Engineering,  \r{A}bo Akademi University, \r{A}bo, Finland}

%\collaboration{FUNCQC project}%\noaffiliation

\date{\today}% It is always \today, today,
             %  but any date may be explicitly specified

\begin{abstract}
We study classical spin models on the 1/1 Tsai-type approximant lattice using Monte Carlo and mean-field methods. Our aim is to understand whether the phase diagram differences between Gd- and Tb-based approximants can be attributed to anisotropy induced by the crystal-electric field. To address this question, we treat Gd ions as Heisenberg spins and Tb ions as Ising spins. Additionally, we consider the presence of the RKKY interaction to replicate the experimentally observed correlation between magnetic properties and electron concentration. Surprisingly, our findings show that the transition between ferromagnetic and antiferromagnetic order remains unaltered by the anisotropy, even when accounting for the dipole interaction. We conclude that a more comprehensive model, extending beyond the free-electron gas RKKY interaction, is likely required to fully understand the distinctions between Gd- and Tb-based approximants. Our work represents a systematic exploration of the impact of anisotropy on the ground-state properties of classical spin models in quasicrystal approximants.
\end{abstract}

%\keywords{Suggested keywords}%Use showkeys class option if keyword
                              %display desired
\maketitle

%\tableofcontents

\section{\label{sec:level1}Introduction}
Quasicrystals provide a unique setting to theorists and experimentalists alike: as well as having crystallographically forbidden rotational symmetries, quasicrystals possess macroscopic structural order that never repeats\cite{ShechtmanBlech,SteinhardtLevine,LEVINE198585}. As most of solid state physics relies on periodicity, there is considerable difficulty in adapting existing methods to treat quasiperiodic systems. However, a possible stepping stone towards studying quasicrystals presents itself in the form of approximants, which are sequences of periodic materials with increasing unit cell sizes. Locally, their structure is the same as that of the limiting quasicrystal; if the correlation length of the observables of interest is shorter than the size of the unit cell, one would therefore expect the physical properties to be similar. Furthermore, a thorough understanding of the properties of the approximants is necessary to identify what makes the quasiperiodic phase unique. In this paper, we present a theoretical study of classical spin models on a Tsai-type approximant as a baseline for further work on their magnetic properties.

Tsai-type materials\cite{Tsai2000} are a class of complex intermetallics which can occur both in quasicrystalline form, and as periodic approximants. The materials contain at least two different elements\cite{Takakura2007}, and are built from clusters of several concentric polyhedral shells. Of particular interest is the icosahedral shell, which can host the magnetic rare-earth elements. Since the discovery of Tsai-type clusters containing rare-earth elements there has been extensive research into their magnetic properties, with the main focus on finding a quasiperiodic system with long-range magnetic order\cite{Goldman_2014}. A quasicrystalline ferromagnet was found by Tamura et al.\cite{Tamura2021} in a Tsai-type Au-Si-Tb system in 2021, but antiferromagnetic order is yet to be observed in quasicrystals despite antiferromagnetic phases being allowed by symmetry arguments\cite{LifshitzExpect}.

\begin{figure}[htb]
    \centering
    \includegraphics[width = 0.95\linewidth]{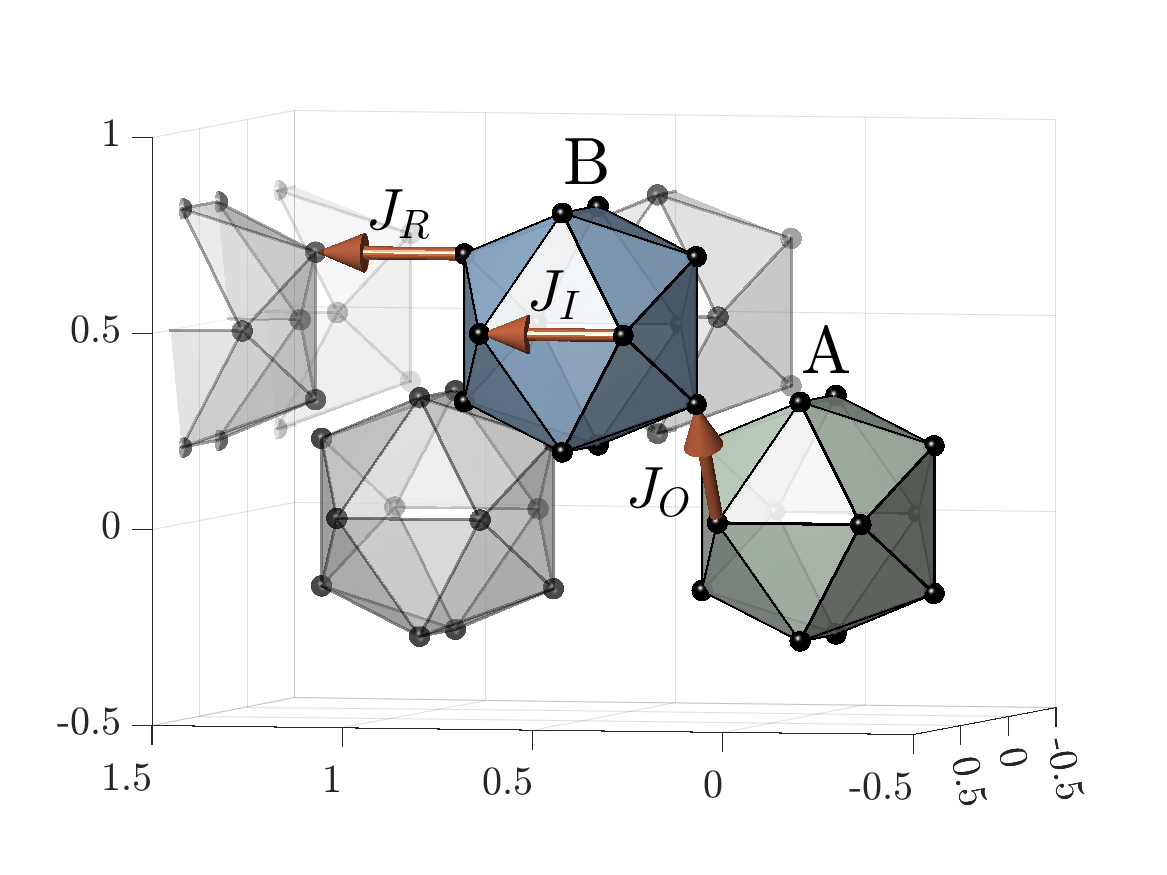}
    \caption{Geometry of the cubic unit cell, in units of the lattice parameter. Rare earth sites occur on the vertices of the icosahedra, which in turn are centered at $(0, 0, 0)$ labeled "A"; and $(1/2, 1/2, 1/2)$, labeled "B", respectively. Clusters from outside the unit cell are greyed out. $J_\text{O}$, $J_\text{I}$ and $J_\text{R}$ denote the three nonequivalent nearest-neighbor distances we consider for the direct exchange terms, see Sec.~\ref{sec:Hamiltonian}}
    \label{fig:geometry}
\end{figure}

As Tsai-type quasicrystals are difficult to investigate from both theoretical and experimental viewpoints, a recurring theme has been to study the periodic approximants to determine which parameter ranges to prioritize in the pursuit of quasiperiodic magnetic order\cite{21approx,Takakura2022}.The approximants are labeled by a rational number approximating the golden ratio, $\tau = (1 + \sqrt{5})/2$. Simplest among the approximants are the 1/1 approximants, which resemble body-centered cubic lattices decorated with Tsai-type clusters: see Fig.~\ref{fig:geometry} for the magnetic lattice. For the rest of this paper, the word approximant is strictly referring to a 1/1 Tsai-type approximant.

\begin{figure}[tb]
    \centering
    \includegraphics[width = 0.95\linewidth]{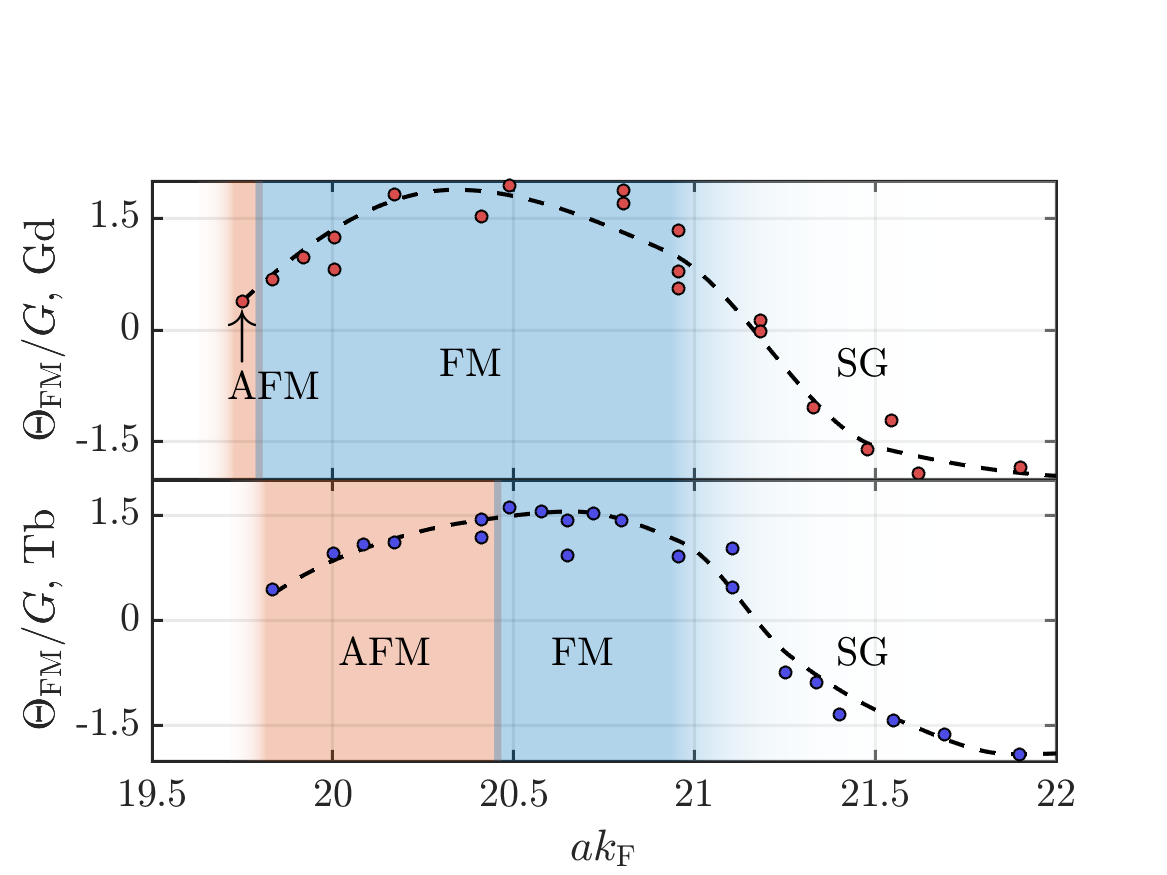}
    \caption{Experimentally measured reduced Curie-Weiss temperatures in Kelvin for Gd-based (top, red dots) and Tb-based (bottom, blue dots) 1/1-approximants versus $a k_\text{F}$, lattice constant $a$ times the Fermi wavevector $k_\text{F}$. Background colours illustrate an approximate magnetic phase diagram. The ordinate values have been divided by the de-Gennes factor, see Eq.~(\ref{eq:rkky_strength}) and the following discussion. Dashed lines are guides to the eye. Adapted from Suzuki et al \cite{Suzuki2021}.}
    \label{fig:suzuki_cw}
\end{figure}

Suzuki et al noted that the Curie-Weiss temperature, which is proportional to the mean magnetic coupling, varies with the conduction electron density for a set of ternary approximant materials \cite{Suzuki2021}. We reproduce their result in Fig.~\ref{fig:suzuki_cw}, where we have modified the abscissa to represent the Fermi wavevector as estimated by Eq.~(\ref{eq:kf_to_nu}). There is a critical value for $k_\text{F}$ below which Tb-based approximant samples transition from ferromagnetic (FM) to antiferromagnetic (AFM)\cite{Hiroto_2014}. However, Gd-based approximants favor FM order, and their transition point is at much lower electron concentration than their Tb counterparts in spite of the similarities in Curie-Weiss temperatures. The leading hypothesis is that this is due to the difference in on-site anisotropies experienced by Gd- and Tb-ions\cite{Hiroto_2020}.

The simplest way to model the conduction electrons is to treat them as a Fermi gas with a spherical Fermi surface which is coupled to the rare-earth magnetic moments through an antiferromagnetic exchange term. This gives rise to an effective coupling between the rare-earth moments: the Ruderman-Kittel-Kasuya-Yosida (RKKY) interaction\cite{RudermanKittel,Kasuya,Yosida}. Sugimoto et al.\cite{Sugimoto2016} presented a phenomenological model of approximants based on the RKKY interaction and easy-plane anisotropy, which reproduces magnetic states observed in neutron diffraction experiments on Tb-Au-Si approximants. Miyazaki et al.\cite{Miyazaki2019} investigated classical Heisenberg spins interacting through RKKY; the resulting ground-state phase diagram as a function of electron concentration correlates with that found in experiments.

In this work, we study classical spin models on the Tsai-type 1/1 approximant lattice using a Monte Carlo approach. We investigate the magnetic properties under the RKKY interaction, but we also consider direct exchange and dipole-dipole interactions in a systematic fashion. In order to connect the simulations to neutron diffraction results, focus lies on the ground state properties of each class of interaction as well as on investigating how the phase diagram under the RKKY interaction is affected by including the direct exchange and dipole interactions. Furthermore, we introduce an Ising ansatz to model the crystal field effect on Tb to determine if strong anisotropy has an effect on the magnetic properties. This model has easy-axis anisotropy, and we contrast it against the case of the isotropic Heisenberg spin symmetry as a model for Gd. Understanding the difference between Tb- and Gd-based approximants could be important for synthesizing magnetic quasicrystals, as approximants have a greater parameter range where they are thermodynamically stable compared to their quasicrystalline counterpart. Therefore, understanding how different rare-earth ions behave in Tsai-type clusters could be important for engineering the quasicrystal magnetic properties. \cite{Goldman_2014}

The paper is organized as follows: in Sec.~\ref{sec:model}, we introduce our model, starting with the lattice geometry, followed by a discussion of the easy-axis model we employ for the crystal electric field, before ending on a presentation of the Hamiltonians under study. In Sec.~\ref{sec:method}, we describe our Monte Carlo approach.  The results are presented and discussed in Sec.~\ref{sec:results}, where we begin by determining the magnetic properties of the direct exchange, RKKY and dipole terms separately. Finally, we consider how dipole and direct exchange terms influence the RKKY ground states as the Fermi wavevector is changed. 

\section{Model}\label{sec:model}

\subsection{Lattice}
If rare-earth ions are present in the Tsai-type 1/1 approximant, they are located at the positions shown as vertices in Fig.~\ref{fig:geometry}. The magnetic lattice is a body-centered cubic (BCC) arrangement of clusters of rare-earth ions, which are in turn located on the vertices of icosahedra. The vertices of an icosahedron can be represented by the cyclic permutations of the position vector $\mathbf{r}_i \propto (\pm 1 \pm\tau ~ 0)$, where $\tau = (1 + \sqrt{5})/2$ is the golden ratio. The cubic unit cell parameter, $a$, is approximately $15$Å. In this work we take the radius of the icosahedral clusters to be $r_\text{Ico} \approx 0.365a$.

The full decoration of the BCC unit cell is made up of a series of concentric shells; an illustration is provided by Suzuki et al \cite{Suzuki2021}. The icosahedra containing rare-earth ions are surrounded by shells containing Au and Si. For our purpose, the key property is this: certain lattice sites are partially occupied by either Au or Si, and this occupancy can be tuned by sample composition. As Au and Si have different valency, one can synthesize materials with different conduction electron density, which we model with a corresponding Fermi wavevector. In spite of the disordered nature of the materials we aim to study, we take the icosahedra to be perfectly symmetric and of the same radius throughout the lattice.

\subsection{Spin symmetry}\label{subsec:spinsym}
The crystal electric field in real materials typically induces a magnetic anisotropy depending on the orbital states of the electrons in the unfilled ionic shells. Thus, the local magnetic moments have a set of preferred directions. In this study, our starting point is to neglect the anisotropy and model the ionic magnetic moments as classical Heisenberg spins. This model should approximate the behavior of Gd-based compounds: by Hund's rules, the electronic ground state of a free Gd ion has zero orbital angular momentum and its 4$f$-electron density is spherically symmetric; therefore the crystal electric field will have little effect on its magnetic properties\cite{coey_2010}. 

\begin{figure}[ht]
    \centering
    \includegraphics[width = 0.95\linewidth]{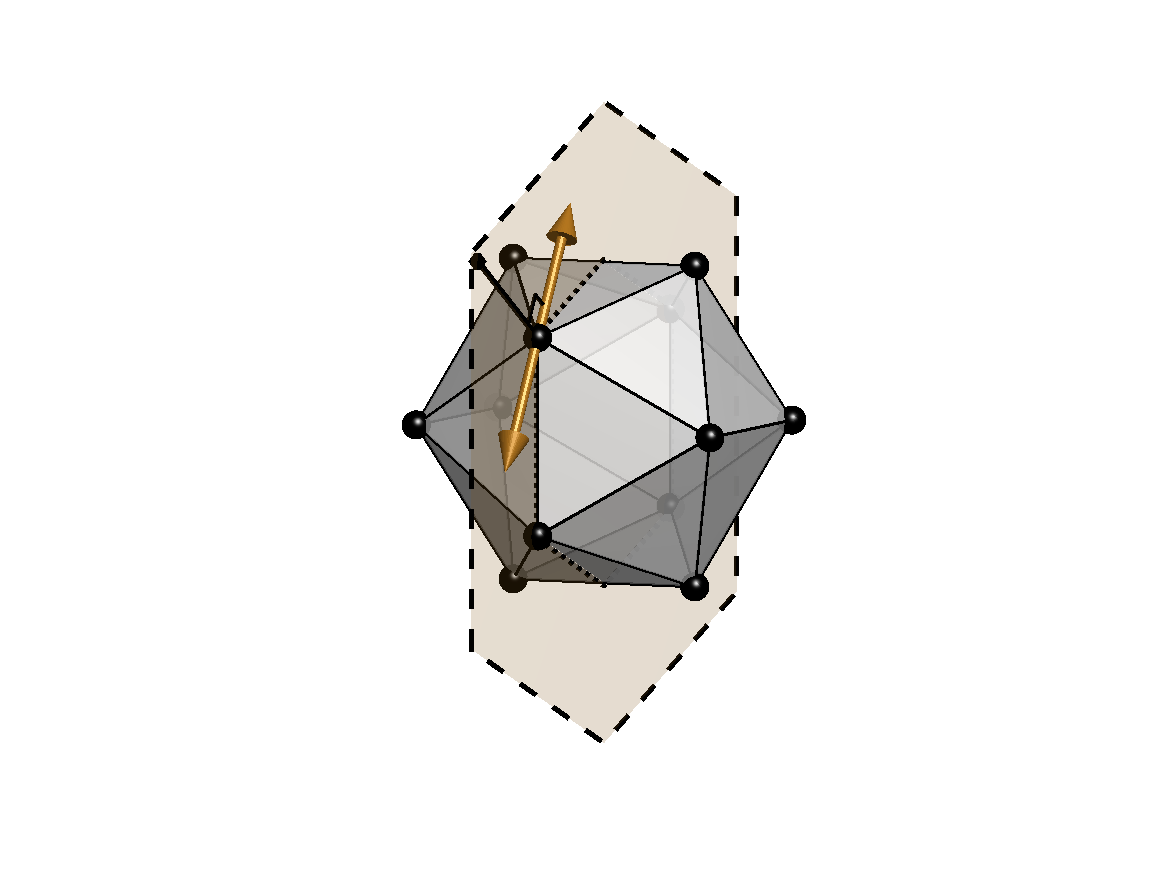}
    \caption{ Spin symmetry of the Ising kind. Transparent hexagon denotes a mirror plane of the lattice, thick black line the normal vector of the icosahedral cluster, and orange vectors show the two possible choices of directions for the Ising spin $\mathbf{S}_i$.}
    \label{fig:ising_anisotropy}
\end{figure}

To study the hypopthesis that the main difference between the Tb and Gd approximant phase diagrams is due to on-site anisotropy, we model the Tb-moments as Ising spins pointing along the icosahedral cluster normal, and inside a mirror plane of the lattice. For example, we assume the site at $\mathbf{r} \propto (\tau 1 0)$ has two possible choices for the Ising spin $\mathbf{S}_i \propto \pm (-1 \tau 0)$, see Fig.~\ref{fig:ising_anisotropy}. We neglect any angular deviation from the idealized case discussed above as the point is not to match the actual anisotropy present in the material, but to put our model in as stark a contrast to the Heisenberg symmetry as possible to maximize the difference in the observed phase diagram. We note, however, that magnetic structures close to those that are acquired from our ansatz have been derived as possible solutions in point charge model calculations \cite{Watanabe2021}.

\subsection{The Hamiltonian}\label{sec:Hamiltonian}

We study the thermodynamic properties of classical Heisenberg and Ising model spins interacting via the Hamiltonian
\begin{equation}\label{eq:Hamiltonian_general}
    \hamil = \hamil_{\text{DE}} + \hamil_{\text{RKKY}} + \hamil_{\text{Dipole}}.
\end{equation}
The class of terms contained in $\mathcal{H}_\text{DE}$ are short-range direct exchange (DE) interactions of the exchange form,
\begin{equation}\label{eq:Hamiltonian_nn}
    \mathcal{H}_\text{DE} = -\sum_{i, j} J_{i j} \mathbf{S}_i \cdot \mathbf{S}_j;
\end{equation}
where $J_{i j} = 0$ if $|\mathbf{r}_{i,j}|\geq 0.39 a$, and otherwise takes the values $J_\text{O}$ for $|\mathbf{r}_{i,j}| \approx 0.360a$, $J_\text{I}$  for $|\mathbf{r}_{i,j}| \approx 0.378a$, or $J_\text{R}$  for $|\mathbf{r}_{i,j}| \approx 0.392a$ according to the relative placement of sites $i$ and $j$ according to Fig.~\ref{fig:geometry}. These three shortest distances are within $\sim10\%$ of one another, whereas the next shortest distance is approximately $0.532a$, and we neglect the corresponding direct exchange term. Given the bipartite nature of the lattice, the key geometrical observation is the following: only $J_\text{O}$ acts between sublattices A and B, whereas both $J_\text{I}$ and $J_\text{R}$ act within the same sublattice.

The RKKY interaction originates from the indirect exchange mediated by the conduction electrons, and is of the form\cite{RudermanKittel,Kasuya,Yosida}
\begin{equation}\label{eq:Hamiltonian_rkky}
    \hamil_{\text{RKKY}} = -J_{\text{RKKY}} \sum_{|\mathbf{r}_{i,j}| < r_\text{c}}(2 k_\text{F} r_\text{Ico})^3 f(2 k_{\text F} |\mathbf{r}_{i,j}|) \mathbf{S}_i \cdot \mathbf{S}_j,
\end{equation}
where $r_\text{c} = 5a$ is a cutoff radius, $k_{\text{F}}$ is the Fermi wavevector and $f(r)$ is given by
\begin{equation}\label{eq:friedel}
    f(r) = \frac{\sin(r) - r\cos(r)}{r^4}.
\end{equation}
The presence of the cutoff of the RKKY interaction is motivated by disorder in the lattice: the effective range of the interaction is limited by the conduction electron mean-free path. As the Fermi wavevector is changed through doping, $r_\text{c}$ should in principle correlate with $k_F$. However, we neglect this relation and instead choose a cutoff radius $r_\text{c} = 5 a$, as we find convergence of our coupling constants at this point.

The relative strength of the RKKY interaction depends on the distance between the spins, the electron density and the rare-earth element: 
\begin{equation}\label{eq:rkky_strength} %TODISH: is the nu the same as in kf_to_nu? If not, the magnitude estimate improves significantly if it's the number of electrons in the 4f-shell!
    J_\text{RKKY} = \frac{9 \pi m_\text{e} J_{sf}^2 \nu^2}{32 \hbar^2 k_\text{F}^2} \frac{1}{(2 k_\text{F} r_\text{Ico})^3} G
\end{equation}
where $\nu$ is the number of valence electrons per atom and $G = (g - 1)^2 J(J + 1)$ is the de Gennes factor for total angular momentum modulus $J$ (in units of $\hbar$). Neglecting partial occupancies, the total number of atoms per cubic unit cell is $N_\text{a} \approx 176$ \cite{PseudoTsai}, and the Fermi wavevector can be related to the electron-per-atom ratio $\nu$ as \cite{Kittel2013}
\begin{equation}\label{eq:kf_to_nu}
    k_\text{F} = (3 \pi^2 N_\text{a} \nu)^{1/3}/a.
\end{equation}
In order to investigate the experimentally relevant interval $\nu \in [1, 2.2]$, we study the corresponding interval for the dimensionless parameter $a k_\text{F} \in [17, 22.5]$. 

The dipolar interaction is defined by
\begin{equation}\label{eq:Hamiltonian_dipole}
    \hamil_{\text{Dipole}} = -J_{\text{D}} \sum_{i,j}r_{\text{Ico}}^3\frac{3 (\mathbf{S}_i \cdot \hat{r}_{i,j})(\mathbf{S}_j\cdot \hat{r}_{i,j}) - \mathbf{S}_i\cdot \mathbf{S}_j}{|\mathbf{r}_{i,j}|^3},
\end{equation}
where $\hat{r}_{i,j}$ is the relative unit vector between sites $i$ and $j$. Assuming free rare-earth ions, $J_\text{D}$ is given by\cite{coey_2010}
\begin{equation}
    J_{\text{D}} = \frac{\mu_0}{4 \pi} \frac{\mu_\text{B} ^ 2}{r_{\text{Ico}}^3} (g J)^2.
\end{equation}
%TODO: Note that the dipole coupling strength is on the order of $5\%$ of the experimental phase transition temperatures typically observed in the approximants\cite{TakaChiral,Tamura2021}, and although the dipolar interaction is often omitted it may nevertheless be of a sufficient magnitude to impact the magnetic properties of the approximant materials.
Inserting values for Gd, we find $J_\text{D}\approx 0.2~\mathrm{K}$, and for Tb, $J_\text{D}\approx0.3$ K.  Taking the typical value for the exchange parameter $J_{sf} \sim 0.2$eV in Eq.~(\ref{eq:rkky_strength})\cite{coey_2010}, and $m_\text{e}$ taken to be the free electron mass, the estimated magnitude for $J_\text{RKKY}$ in Gd-based approximants is found to be $\sim 0.3$K --- of comparable magnitude to $J_\text{D}$. Thus, we stress that we cannot neglect the dipole interaction \textit{a priori}, as is usually done for e.g. room-temperature ferromagnets.\cite{ashcroft1976solid}

\section{Methods}\label{sec:method}
In this section we introduce the methods we use to analyze the properties of the terms in Eq.~(\ref{eq:Hamiltonian_general}). To begin with, we reproduce the mean-field theory description of a bipartite Heisenberg model in Subsec.~\ref{subsec:mft}. The purpose of this is twofold: first, it connects to the recent experimental focus on determining the Curie-Weiss temperature; and second, it provides an intuitively clear framework for explaining the transition between ferro- and antiferromagnetic order we observe in the Monte Carlo simulations, and in particular the role of the effective inter-sublattice coupling $u_\text{AB}$.
\subsection{Mean-field theory}\label{subsec:mft}
Consider a general Hamiltonian with scalar couplings,
\begin{equation}
    \hamil = - \sum_{ij} J_{i j} \mathbf{S}_i \cdot \mathbf{S}_j.
\end{equation}
In the framework of mean-field theory one neglects the correlations between spins, and each site interacts magnetically with the average field generated by the others. For Heisenberg spins, each site in Fig.~\ref{fig:geometry} is equivalent by symmetry. Thus, one can make the ansatz that the $z$-components of spins on the respective sublattices all have the same mean value: $\braket{\mathbf{S}^z_{i \in \text{A}}} = s_\text{A}$  and $\braket{\mathbf{S}^z_{i \in \text{B}}} = s_\text{B}$. The self-consistency condition for a spin at site $i$ in the $\text{A}$-sublattice becomes
\begin{equation}\label{eq:self_consistency}
    s_\text{A} = \frac{\text{Tr}_i[s_i e^{\beta (u_\text{AA} s_\text{A} + u_\text{AB} s_\text{B} + h_\text{ext})s_i}]}{\text{Tr}_i[e^{\beta (u_\text{AA} s_\text{A} + u_\text{AB} s_\text{B} + h_\text{ext})s_i}]},
\end{equation}
where $\text{Tr}_i (\cdot) = \int_{-1}^{+1} 2 \pi ds_i (\cdot)$ is an integral over the possible orientations of $\mathbf{S}_{i}$, and the mean-field couplings are given by
\begin{equation}
    u_\text{AA} = \sum_{j \in \text{A}} J_{ij}, \quad i \in \text{A},
\end{equation}
describing the effective coupling within sublattices, and 
\begin{equation}
    u_\text{AB} = \sum_{j \in \text{B}} J_{ij}, \quad i \in \text{A}.
\end{equation}
describing the effective coupling between sublattices. There are analogous expressions for $i \in B$. By symmetry, $u_\text{BB} = u_\text{AA}$ and $u_\text{BA} = u_\text{AB}$. We use the letter $u$ here as opposed to the customary $J$ to distinguish the mean-field effective couplings from the model parameters in the Hamiltonian.

Computing the traces in Eq.~(\ref{eq:self_consistency}) and expanding to linear order in $\beta$, one arrives at a matrix system for the average configurations
\begin{equation}\label{eq:matrix_consistency}
    \left[ \begin{array}{cc}
        u_\text{AA} - 3/\beta &  u_\text{AB} \\
        u_\text{BA} &  u_\text{BB} - 3/\beta
    \end{array} \right] 
    \left[ \begin{array}{c}
        {s_\text{A}} \\
        {s_\text{B}}
    \end{array} \right] = h_\text{ext} \left[ \begin{array}{c}
        1 \\
        \sigma
    \end{array} \right],
\end{equation}
where $\sigma = +1$ ($\sigma = -1$) probes for FM (AFM) order, see Eq.~(\ref{eq:stagmag}). Inverting the matrix and differentiating the resulting magnetization with respect to $h_\text{ext}$, one arrives at
\begin{equation}
    \chi_{zz} \sim \frac{1}{1 - \beta (u_\text{AA} + u_\text{AB})/3},
\end{equation}
corresponding to $\sigma = +1$ in Eq.~(\ref{eq:matrix_consistency}), and a staggered susceptibility
\begin{equation}
    \chi_s \sim \frac{1}{1 - \beta (u_\text{AA} - u_\text{AB})/3},
\end{equation}
corresponding to $\sigma = -1$. In both cases, the critical temperature is given by 
\begin{equation}\label{eq:tc_mft}
    T_\text{c} \sim \frac{u_\text{AA} + |u_\text{AB}|}{3}.
\end{equation}
A property of this mean-field solution that turns out to be important is the following: the sign of $u_\text{AB}$ indicates whether one can expect a ferromagnet or an antiferromagnet, which we use to discuss the magnetic phase diagrams in Sec.~\ref{sec:results}. \cite{coey_2010}

\subsection{Numerical methods}\label{subsec:mc}
We use two different Monte Carlo approaches: the Metropolis algorithm\cite{RubinsteinKroese} for Ising spins and the heat-bath\cite{Miyatake1986} algorithm for Heisenberg spins. Unless otherwise noted, we simulate a system size of $24\cdot4^3= 1536$ sites, i.e. a linear system size of $L = 4$ cubic unit cells. Furthermore, we employ periodic boundary conditions. In order to determine the ground state, we employ the simulated annealing approach using $10^6$ updates per site and temperature, and use a uniformly spaced distribution of 10 temperatures.

When simulating dipolar systems using periodic boundary conditions, care has to be taken as the series in Eq.~(\ref{eq:Hamiltonian_dipole}) is in that case conditionally convergent. Any physical sample, however, has a boundary whose shape will influence the magnetostatic energy. The usual way to implement the boundary is through a shape-dependent Ewald sum, which for a spherically shaped sample yields the effective dipole Hamiltonian \cite{Leeuw1980}
\begin{equation}
        \mathcal{H}_\mathrm{D} = -\frac{1}{2} J_\mathrm{D}\sum_{i\neq j} (\mathbf{S}_i \cdot \nabla) (\mathbf{S}_j \cdot \nabla) \psi(\mathbf{r}_{ij}) + \mathcal{H}_{\mathrm{Boundary}},
\end{equation}
where the scalar potential is
\begin{equation}\label{eq:ewald_split}
    \begin{split}
        \psi(\mathbf{r}) = & \sum_{\mathbf{n} \in \mathbb{Z}^3} \frac{\mathrm{erfc}(\alpha |\mathbf{r} + \mathbf{n}|)}{|\mathbf{r} + \mathbf{n}|} \\
                                  & +\sum_{|\mathbf{n}| \neq 0} \frac{1}{\pi |\mathbf{n}|^2} e^{2 \pi i \mathbf{n} \cdot \mathbf{r}} e^{- \pi^2 |\mathbf{n}|^2/\alpha^2},
    \end{split}
\end{equation}
and $\mathcal{H}_\mathrm{Boundary}$ is given by 
\begin{equation}\label{eq:dipole_boundary}
    H_\text{Boundary} = \frac{2 \pi}{3} \frac{J_\text{D}} {L^3} \sum_{i j} \mathbf{S}_i \cdot \mathbf{S}_j.
\end{equation} 
The parameter $\alpha$ sets the convergence rate of the two different series in Eq.~(\ref{eq:ewald_split}): a large $\alpha$ means that the first series requires fewer terms to converge and the second series require more terms, and vice versa. For our simulations, we used $\alpha = (\pi N)^{1/6}/L$ as an approximation to the optimal value for efficient calculation of $\psi$\cite{Jaubert}.

\subsection{Observables under study}
Using the methods described in Subsec.~\ref{subsec:mc}, we calculate the thermal expectation values of a set of observables. To probe for FM order, we define the magnetization $\mathbf{M}$ as
\begin{equation}\label{eq:mag}
    \mathbf{M} = \sum_{i} \mathbf{S}_i,
\end{equation}
and from the variance of the $z$-component of the magnetization we define the differential susceptibility,
\begin{equation}\label{eq:chi}
    \chi_{zz} = \frac{\braket{M_{z} M_{z}} - \braket{M_{z}}\braket{M_{z}}}{N k_\text{B} T}.
\end{equation}

\begin{figure}[tb]
    \centering
    \includegraphics[width = 0.85\linewidth]{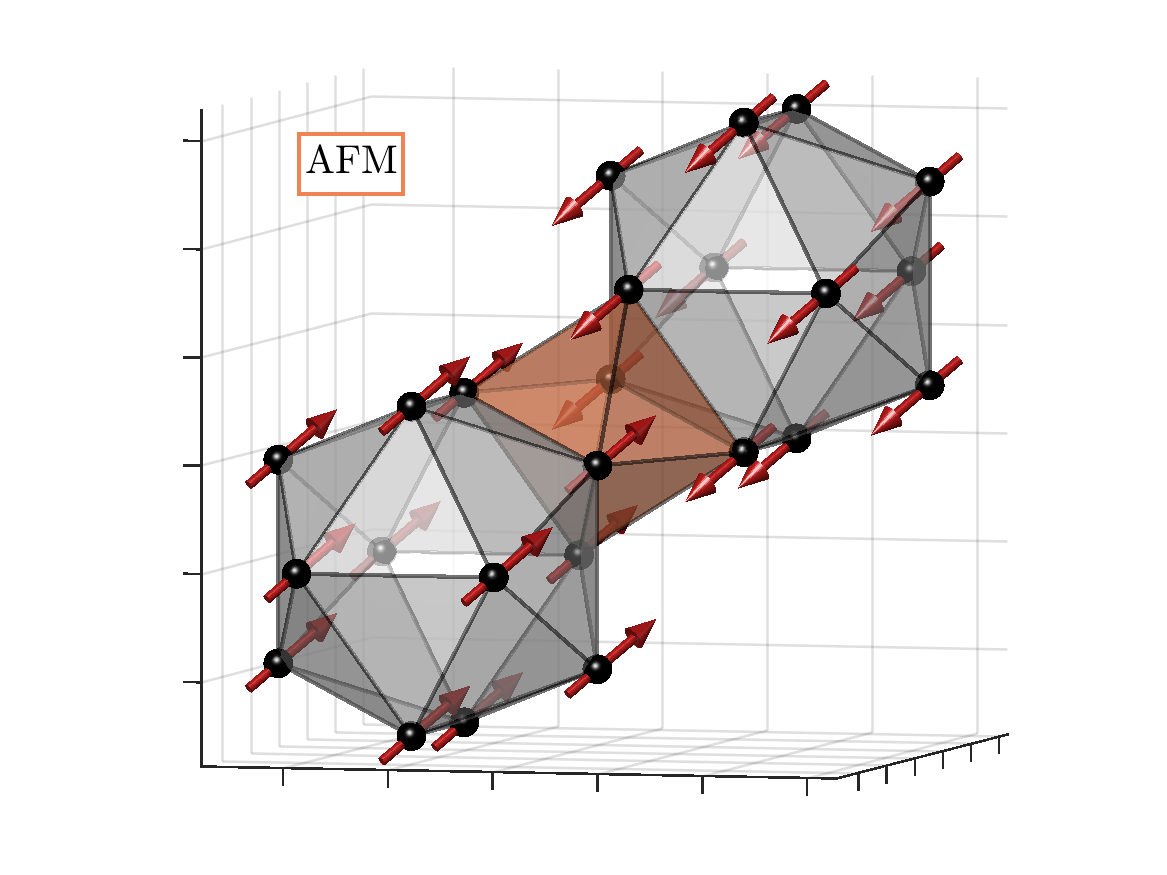}
    \caption{An example AFM state which maximizes the staggered cluster magnetization $|\mathbf{M_s}|$, and therefore also sublattice magnetization. Interstitial octahedron highlighted in orange to illustrate the depth, as the spins are staggered along the 111-direction. }
    \label{fig:afm_example}
\end{figure}

As the lattice is of the BCC class with icosahedral clusters at the BCC sites, one can define a staggered cluster magnetization in analogy with undecorated BCC lattices,
\begin{equation}\label{eq:stagmag}
    \mathbf{M}_\text{s} = \sum_{i \in \text{A}} \mathbf{S}_i - \sum_{i \in \text{B}} \mathbf{S}_i,
\end{equation}
where A and B are sublattices denoted by the corresponding letter in Fig.~\ref{fig:geometry}. An example of a state maximizing $|\mathbf{M}_s|$ is found in Fig.~\ref{fig:afm_example}. Note that a state for which $\braket{|\mathbf{M}_s|} = 0$ could still have antiferromagnetic correlations between neighboring icosahedral clusters. For the purposes of this paper, however, antiferromagnetic order is defined as staggered magnetic order of the form given in Fig.~\ref{fig:afm_example}.

\section{Results and Discussion}\label{sec:results}
In this section, we investigate the differences between the Heisenberg and Ising spin symmetries in the lattice in Fig.~\ref{fig:geometry}. With the numerical method outlined in Sec.~\ref{sec:method}, we consider nearest neighbor, RKKY, and dipole interactions separately in Subsecs.~\ref{subsec:nn}-\ref{subsec:dipole}. Next, we consider the interplay between the RKKY and dipole terms in Subsec.~\ref{subsec:rkky_and_dipole}, as the directional dependence in the dipole interaction might affect the symmetry properties of the ground states. Finally, we modify the inter-sublattice coupling $u_\text{AB}$ by considering $J_\text{O}$ and $J_\text{RKKY}$, and study their mutual effect on the AFM/FM transition.

\subsection{Direct-exchange interactions}\label{subsec:nn}
We begin the presentation of our results by studying the short-range interactions in $\hamil_\text{DE}$, Eq.~(\ref{eq:Hamiltonian_nn}). Although $\hamil_\text{DE}$ does not capture the dependence of the magnetic properties on the electron concentration, its conceptual simplicity allows for direct comparison with mean-field theory. Furthermore, as chemical disorder provides an effective cutoff on the RKKY interaction, with a suitable choice of parameters $\hamil_\text{DE}$ can be seen as a strong disorder limit of $\hamil_\text{RKKY}$.

\begin{figure}[tb]
    \centering
    \includegraphics[width = 0.85\linewidth]{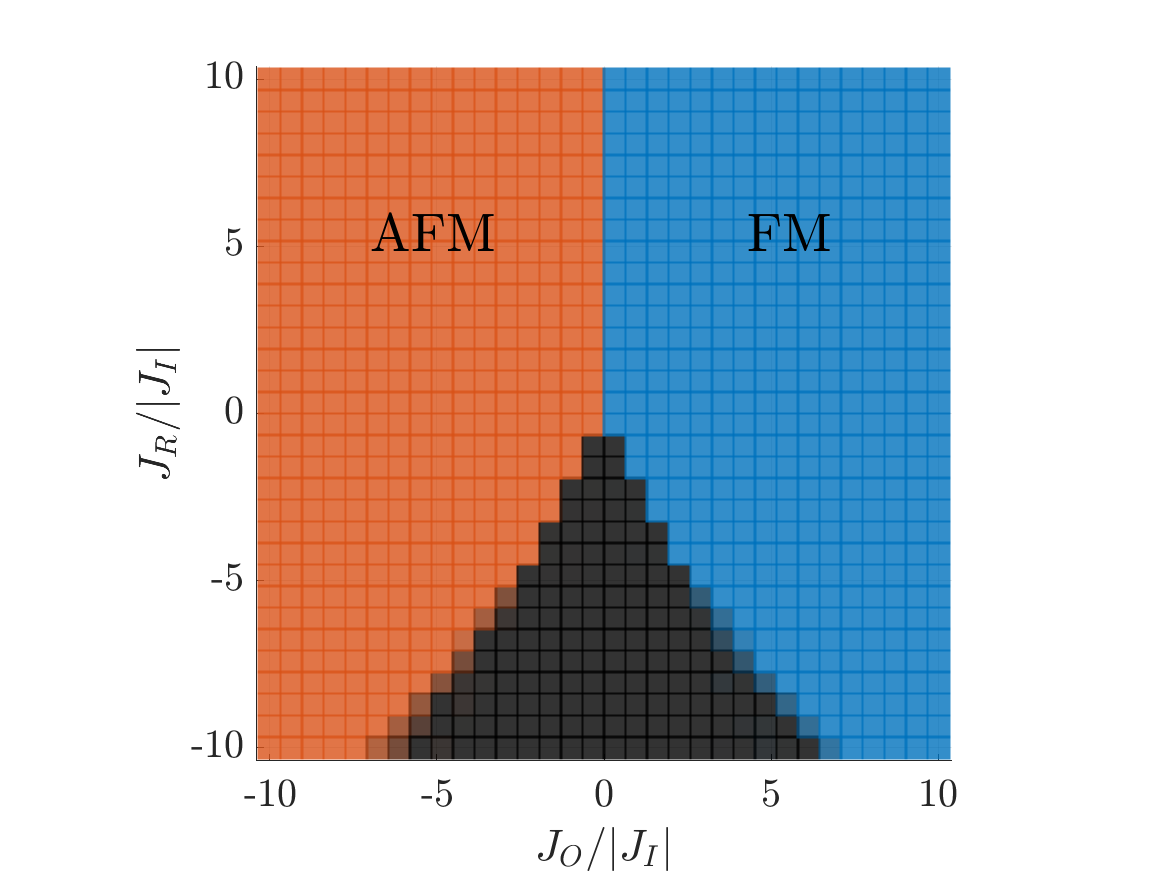}
    \caption{Ground-state phase diagram for Heisenberg spin symmetry under $\mathcal{H}_\text{DE}$ acquired via MC simulation. We have set $J_\text{I} = +1$, and modify the values of $J_\text{O}$ and $J_\text{R}$ along the x- and y-axis, respectively. Orange denotes antiferromagnetic order, and blue ferromagnetic order, as defined by Eqs.~(\ref{eq:mag}) and (\ref{eq:stagmag}). Gray denotes states without a sublattice magnetization.}
    \label{fig:heisen_nn_pm_iplus}
\end{figure}

\begin{figure}[tb]
    \centering
    \includegraphics[width = 0.85\linewidth]{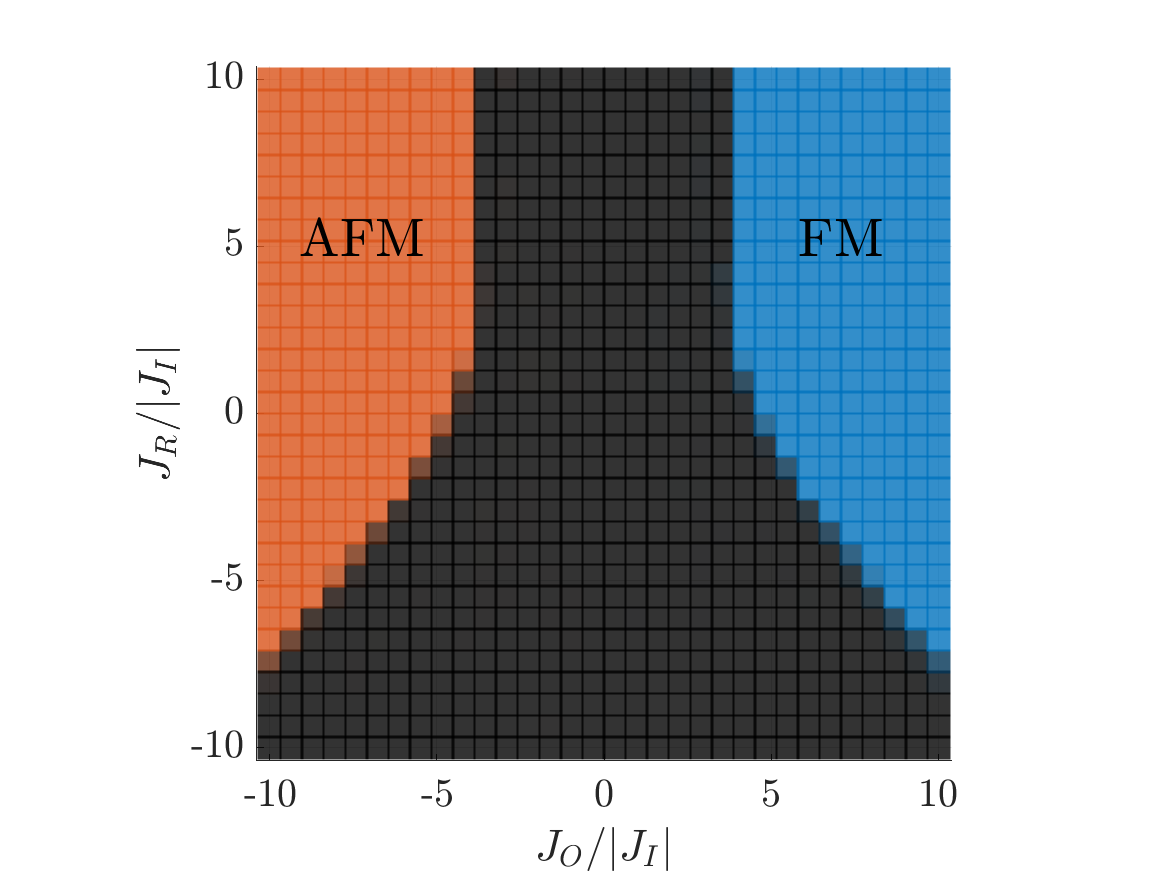}
    \caption{Ground-state phase diagram for Heisenberg spin symmetry under $\mathcal{H}_\text{DE}$ acquired via MC simulation. We have set $J_\text{I} = -1$, and modify the values of $J_\text{O}$ and $J_\text{R}$ along the x- and y-axis, respectively. Orange denotes antiferromagnetic order, and blue ferromagnetic order, as defined by Eqs.~(\ref{eq:mag}) and (\ref{eq:stagmag}). Gray denotes states without a sublattice magnetization.}
    \label{fig:heisen_nn_pm_iminus}
\end{figure}

At zero temperature, only the relative magnitudes of couplings matter for the determination of the magnetic state. Thus, the only parameters determining the ground states of $\mathcal{H}_\text{DE}$ are $J_\text{O}/J_\text{I}$, $J_\text{R}/J_\text{I}$, and the sign of $J_\text{I}$. Therefore, we consider the two cases $J_\text{I} = +1$ and $J_\text{I} = -1$ separately and study the ground state as a function of $J_\text{O}$ and $J_\text{R}$ using Monte Carlo simulation. As seen in Fig.~\ref{fig:geometry}, $J_\text{O}$ sets the strength of the inter-sublattice coupling --- i.e. $u_\text{AB} \sim J_\text{O}$ --- whereas $J_\text{I}$ and $J_\text{R}$ contribute to the intra-sublattice coupling $u_\text{AA}$. By requiring that $T_c > 0$ in Eq.~(\ref{eq:tc_mft}), one can achieve a qualitative description of the phase boundaries of the nearest-neighbor Hamiltonian under Heisenberg symmetry: in the case of $J_\text{I} = +1$ and $J_\text{R} > 0$, Fig.~\ref{fig:heisen_nn_pm_iplus}, the magnetic ordering is determined by the sign of $J_\text{O}$, whereas when $J_\text{R} \lesssim -1$, larger magnitudes of $J_\text{O}$ are needed to observe the AFM/FM ground states in order to overcome the negative $u_\text{AA}$. Similarly, when $J_\text{I} = -1$, the transitions occur at larger $|J_\text{O}|$, see Fig.~\ref{fig:heisen_nn_pm_iminus}. The actual parameter magnitudes at the phase boundaries do not agree with the estimates acquired from Eq.~(\ref{eq:tc_mft}), which is likely due to the fact that the result for $T_c$ is obtained via a high-temperature expansion and the exclusion of states with zero sublattice magnetization from our analysis.

\begin{figure}[tb]
    \centering
    \includegraphics[width = 0.85\linewidth]{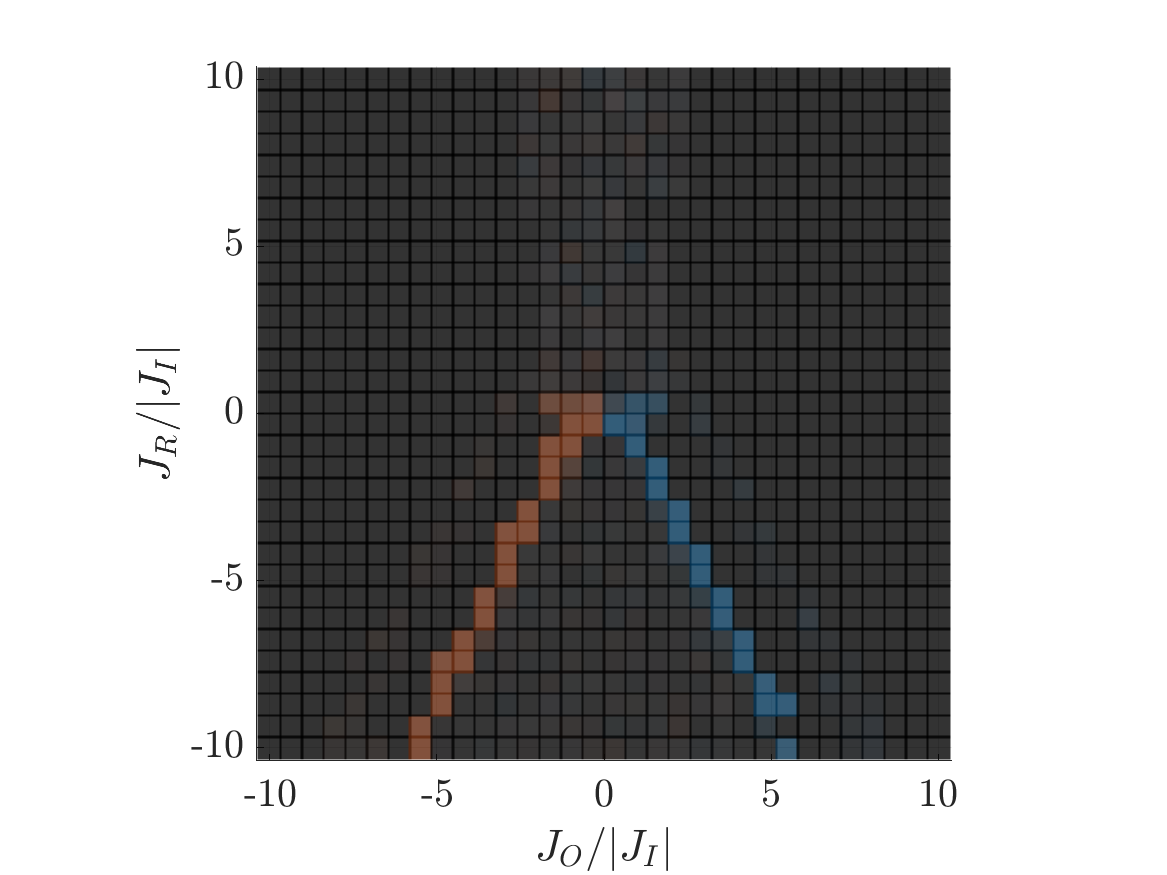}
    \caption{Ground-state phase diagram for Ising spin symmetry under $\mathcal{H}_\text{DE}$ acquired via MC simulation. We have set $J_\text{I} = +1$, and modify the values of $J_\text{O}$ and $J_\text{R}$ along the x- and y-axis, respectively. Orange denotes antiferromagnetic order, and blue ferromagnetic order, as defined by Eqs.~(\ref{eq:mag}) and (\ref{eq:stagmag}). Gray denotes states without a sublattice magnetization. The axis of symmery given by $J_\mathrm{O} = 0$ persists as in Fig.~\ref{fig:heisen_nn_pm_iplus}, but magnetized states are only found in a $\Lambda$-shaped region.}
    \label{fig:ising_nn_pm_iplus}
\end{figure}

The Ising-spin symmetry appears not to have finite values for the magnetization/staggered magnetization except when fine tuning the couplings: see Fig.~\ref{fig:ising_nn_pm_iplus}, where we fix $J_\text{I} = +1$. We find magnetic order defined by Eqs.~(\ref{eq:mag}) and (\ref{eq:stagmag}) only within thin lines in the regime where $J_\text{R} \leq 0$. It appears that the short-range interactions favors ground states without sublattice magntization, except for in the narrow $\Lambda$-shaped region in Fig.~\ref{fig:ising_nn_pm_iplus}. There is a marked contrast between the behaviors of the easy-axis spin system we propose here to the easy-plane one proposed by Sugimoto et al \cite{Sugimoto2016}, the latter showing sublattice magnetization when considering the RKKY interaction truncated to nearest-neighbors without fine-tuning the parameters as seen necessary in Fig.~\ref{fig:ising_nn_pm_iplus}. In order to fully understand the phase diagram, one would need to extend the parameters under study, e.g. consider spin-spin correlation functions --- a task beyond the scope of this paper. 

\subsection{Pure RKKY interaction}\label{subsec:rkky}
\begin{figure}[tb]
    \centering
    \includegraphics[width = 0.95\linewidth]{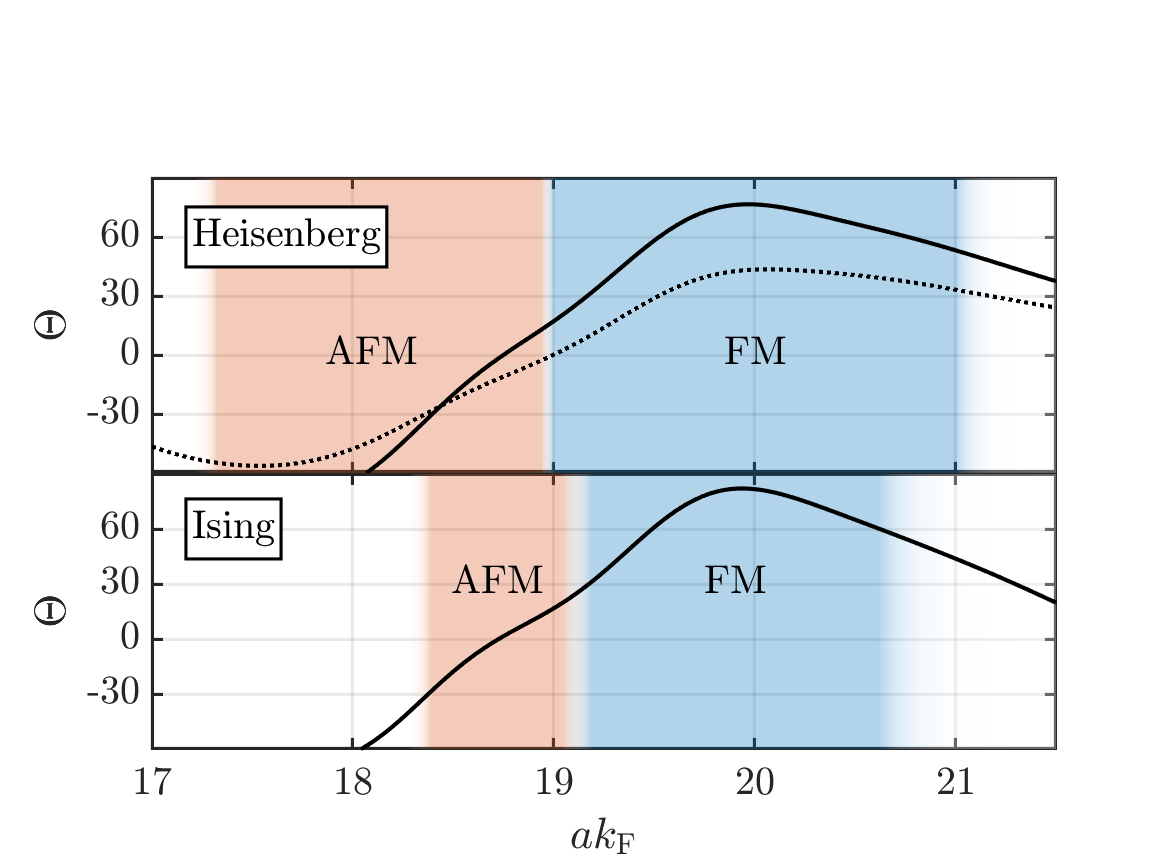}
    \caption{Pure RKKY phase diagram. Orange (blue) color highlights the phase boundaries of the AFM (FM) states. Black lines and dotted line show $\Theta$ and $u_\text{AB}$, respectively --- the latter only defined for Heisenberg symmetry. Regions in white are without sublattice magnetization.}
    \label{fig:mag_plot_rkky_dom}
\end{figure}

When the Hamiltonian contains the RKKY-terms in Eq.~(\ref{eq:Hamiltonian_rkky}) only, the ground state configuration acquired from Monte Carlo simulation correlates with the sign of $u_\text{AB}$ at a given Fermi wavevector, but not with the Curie-Weiss temperature $(u_\text{AB} + u_\text{AA})/3$. That is, it is only the inter-sublattice component of the Curie-Weiss temperature that is an indicator for the magnetic ground state. In Fig.~\ref{fig:mag_plot_rkky_dom}, we show the phase boundaries for the Heisenberg spin symmetries overlapped with $u_\text{AB}$ and the Curie-Weiss temperature as a function of $k_F$. However, as in the case of the direct-exchange Hamiltonian, Eq.~(\ref{eq:Hamiltonian_nn}), the mean-field description only manages to capture the transitions between states with sublattice magnetization. Similar to the experimental situation for Tb- and Gd-based approximants, we find FM and AFM phases for both Ising and Heisenberg spin symmetries under RKKY. 

\begin{figure}[tb]
    \centering
    \includegraphics[width = 0.85\linewidth]{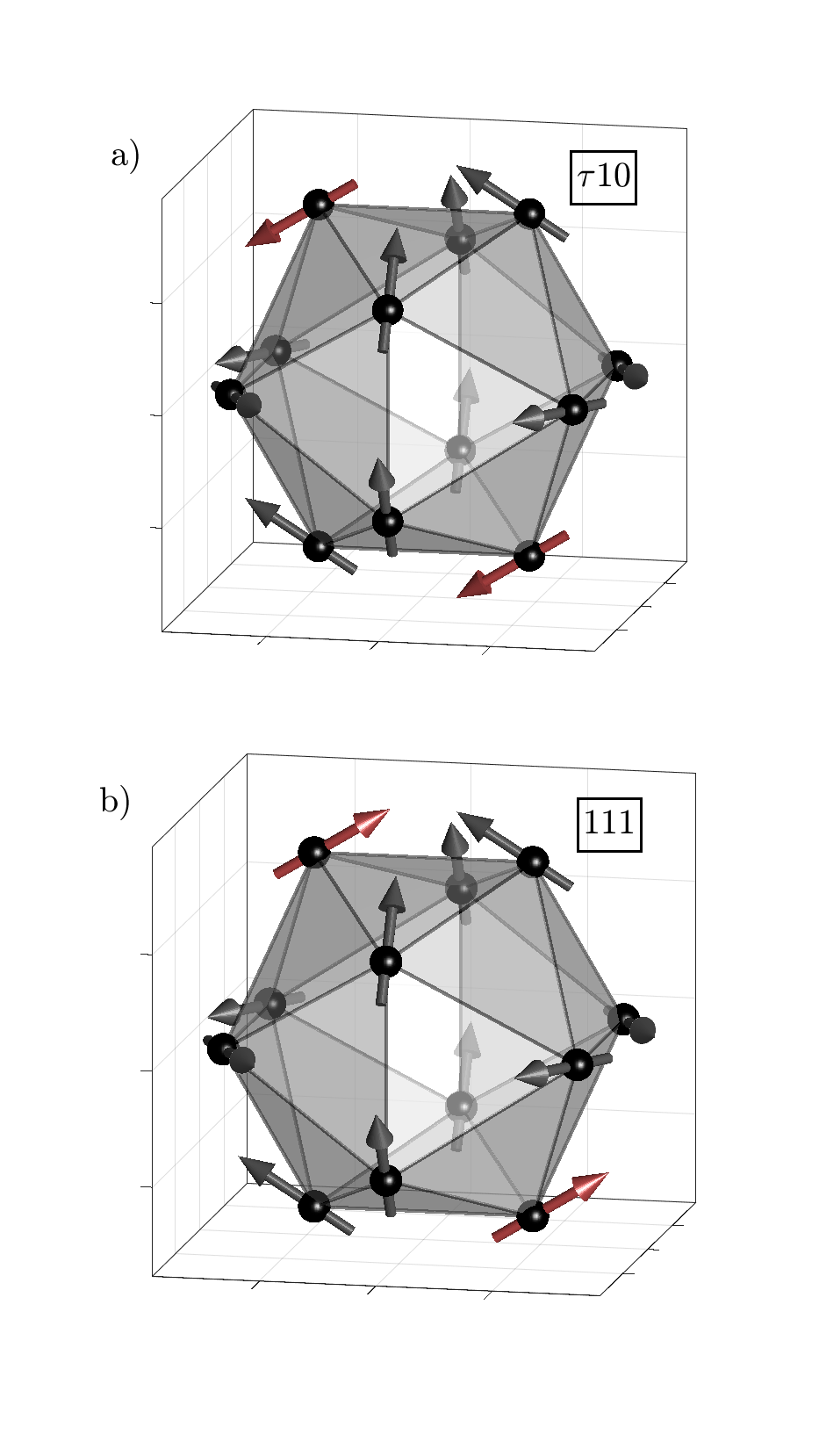}
    \caption{Local cluster orderings in the Ising model system, denoted by the Miller indices of the net magnetization direction: a) the $\tau10$ state b) the $111$ state. The two spins with differing directions between the two states are highlighted in red. We find both antiferromagnetic and ferromagnetic arrangements of these cluster orderings.}
    \label{fig:fmtau10}
\end{figure}

%% SOURCE
As in Miyazaki et al.\cite{Miyazaki2019}, the Heisenberg model reaches its saturation magnetization for all $k_\text{F}$ in the FM regime. However, since both the interactions and the spin symmetry are isotropic, the model cannot select a particular direction. Furthermore, we do not find the incommensurate regime between the FM/AFM phases reported by Miyazaki et al, which could be due to us simulating a smaller system size. In the case of the Ising model, the magnetization saturates at about $53\%$ of the Heisenberg case, see the ferromagnetic ground state in Fig.~\ref{fig:fmtau10}. The possible axes of magnetization for the Ising model ground states are found to point toward the twelve vertices of an icosahedron. Thus, we label its FM (AFM) ground states as FM $\tau 1 0$ (AFM $\tau 1 0$). We note that the symmetry operations relating these directions are not related to the cubic symmetry of the lattice. However, adding the dipole interaction changes this, see Sec.~\ref{subsec:rkky_and_dipole}.

For smaller $k_\text{F}$, both the Heisenberg and Ising models show antiferromagnetic order which is staggered along the 111-direction. The antiferromagnetic states are related to their ferromagnetic counterparts through inverting the spins located on the $B$ clusters in the cubic unit cell, see Fig.~\ref{fig:geometry}. The position of the AFM/FM transition is approximately equal in both of the spin symmetries, which could be due to the long-range interactions coupling sites where the easy-axis vectors are parallel; e.g., the antipodal points in the icosahedral cluster. Thus, those correlations mimic more closely the ground states found in the Heisenberg symmetry case. We discuss this further in Subsec.~\ref{subsec:kf_transition}.

\subsection{Pure dipole interaction}\label{subsec:dipole}

\begin{figure}[tb]
    \centering
    \includegraphics[width=\linewidth]{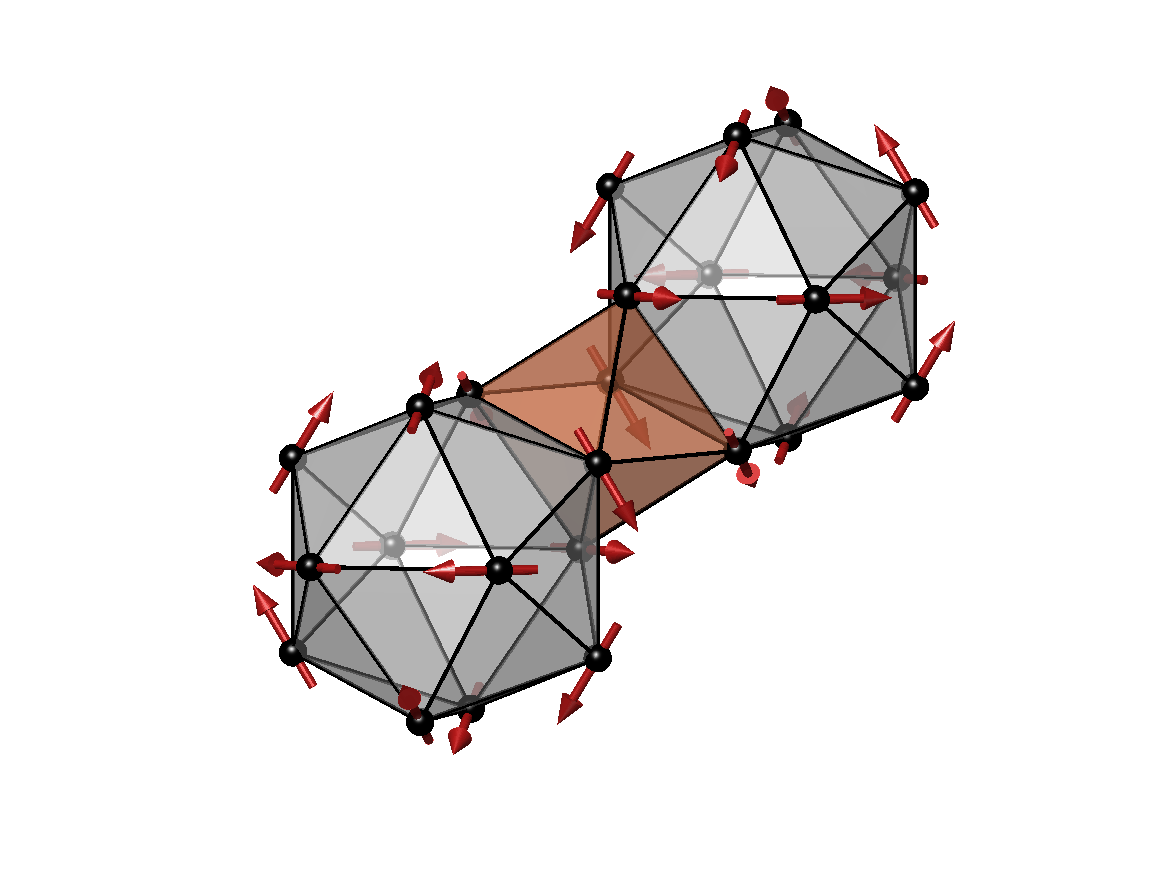}
    \caption{Dipole interaction ground state under Ising symmetry. Note the chirality displayed by the clusters along the 111-direction. }
    \label{fig:dipole_gs_ising}
\end{figure}

\begin{figure}[tb]
    \centering
    \includegraphics[width=\linewidth]{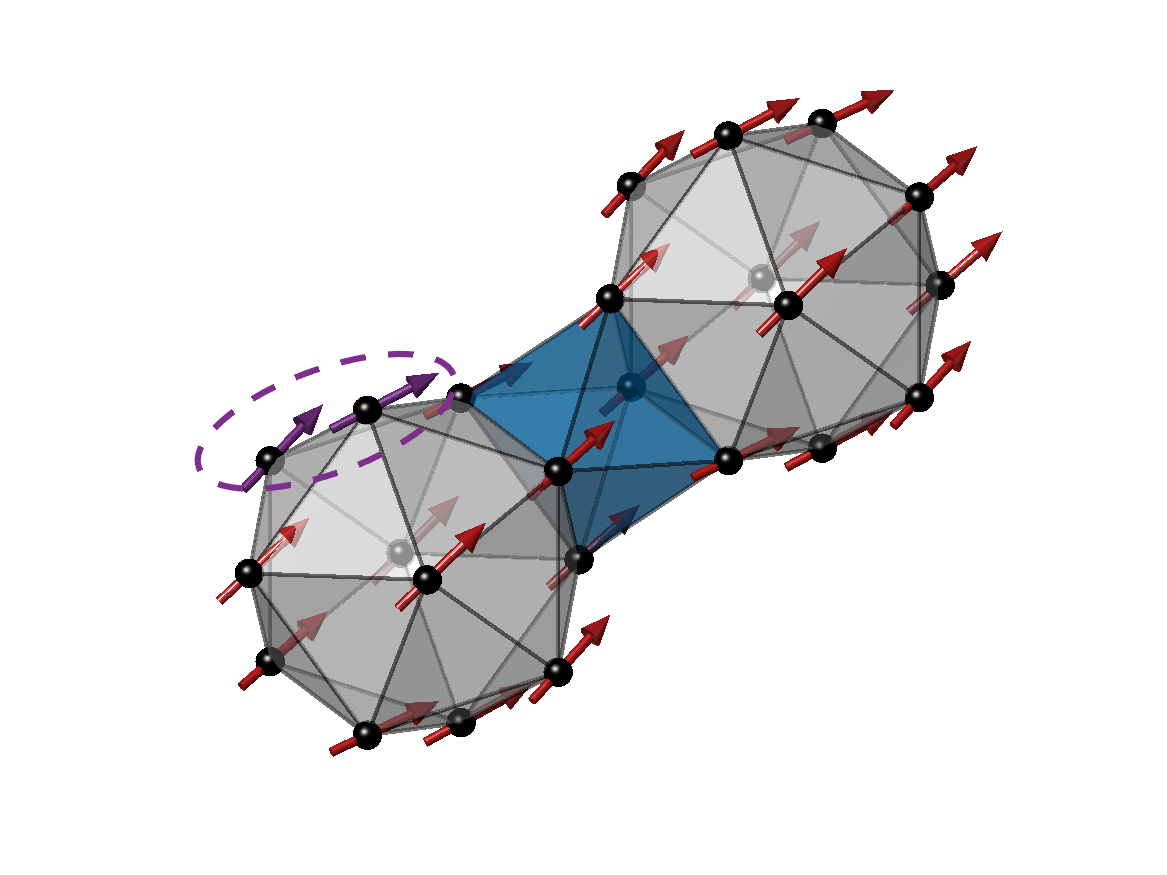}
    \caption{Heisenberg symmetry dipole interaction ground state, neglecting the boundary term. We bring attention to the fact that the structure isn't perfectly collinear, see e.g. neighboring spins highlighted by the dashed purple ellipse.}
    \label{fig:dipole_gs_heisen}
\end{figure}

First, we consider the ground state of the dipole Hamiltonian described by Eq.~(\ref{eq:Hamiltonian_dipole}) with the boundary term in Eq.~(\ref{eq:dipole_boundary}) subtracted, yielding so-called Ewald boundary conditions. This subtraction allows us to study the magnetic structure within the magnetic domains\cite{TwengstromDemag}. We find that Heisenberg spin symmetry yields almost collinear ferromagnetic order in the 111-direction as shown in Fig.~\ref{fig:dipole_gs_heisen}. Note that this is also close to one of the ground states under the RKKY interaction, which means that there may not necessarily be a competition between these two interactions at low temperature. Enforcing Ising symmetry removes the overall magnetization, and the ground state spin configuration features chiral order along the 111 direction, see Fig.~\ref{fig:dipole_gs_ising}. The rotation is clockwise for the A clusters and anticlockwise for the B clusters. %TAKA chiral

\begin{figure}[tb] %TODO: Check x-axis values
    \centering
    \includegraphics[width=\linewidth]{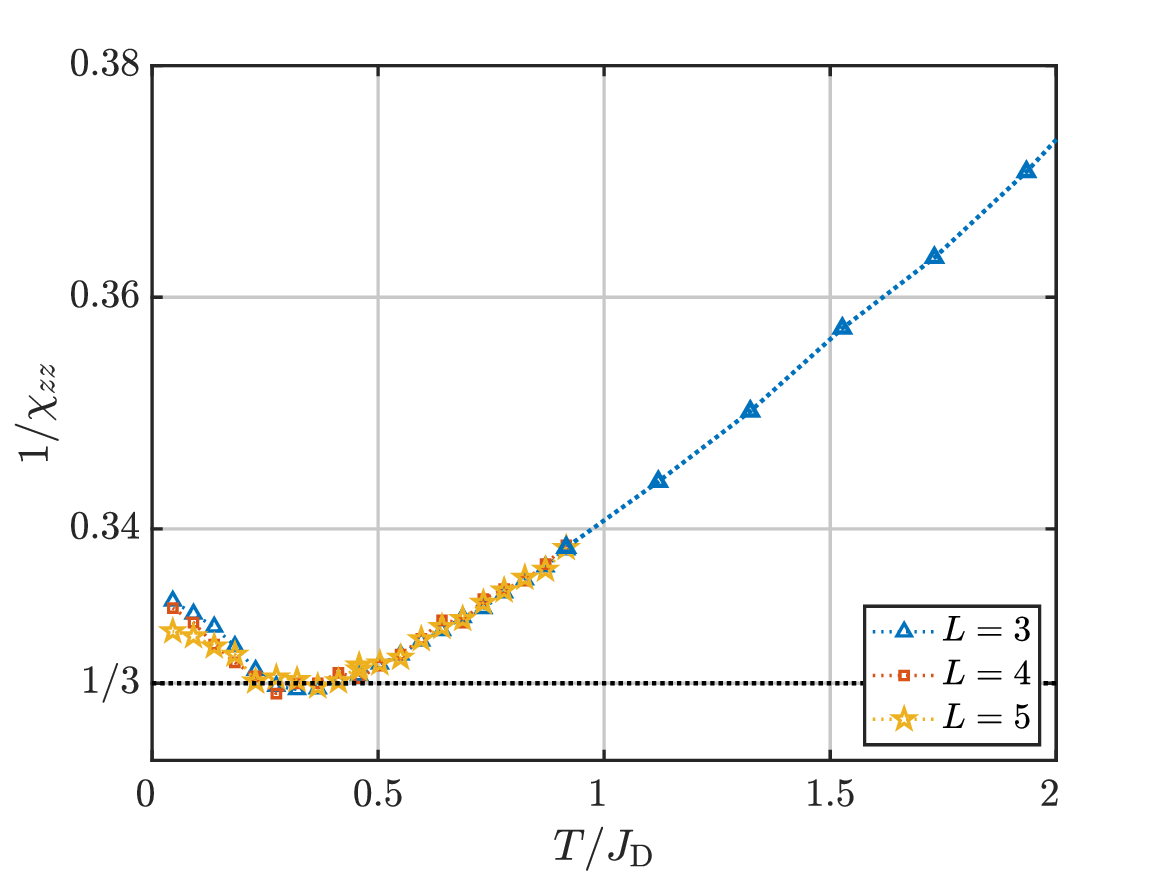}
    \caption{Magnetic susceptibility for the dipole Hamiltonian, Eq.~(\ref{eq:Hamiltonian_dipole}), under Heisenberg symmetry for different linear system sizes $L$. Note the minimum at $1/\chi_{zz} = N = 1/3$, the demagnetizing factor for a spherical sample.} 
    \label{fig:dipole_scaling}
\end{figure}

Since the ground state of the Heisenberg approximant model is a dipolar ferromagnet, we would expect a boundary term to lead to a low-temperature plateau in the susceptibility \cite{TwengstromDemag}.
Indeed, upon reintroducing the boundary term, we observe a minimum in $1/\chi_{zz}$, at $N = 1/3$ for a spherical boundary, but note that the susceptibility rises slightly as the temperature is lowered --- see Fig.~\ref{fig:dipole_scaling}. Our hypothesis is that this behavior is caused by the slight angular deviations from the mean magnetization axis seen in Fig.~\ref{fig:dipole_gs_heisen}.

\subsection{Interplay between RKKY and dipole interactions}\label{subsec:rkky_and_dipole}
As we found in section \ref{subsec:rkky}, anisotropy does not affect the AFM/FM transition point of the approximants under the RKKY interaction. This might be due to the long-range, angularly independent nature of the interaction. Therefore, we investigate the effect of breaking the global rotation symmetry by adding to the RKKY Hamiltonian, Eq.~(\ref{eq:Hamiltonian_rkky}), a dipolar perturbation, Eq.~(\ref{eq:Hamiltonian_dipole}), to see the effect on the phase diagram.

\begin{figure}
    \centering
    \includegraphics[width = 0.95\linewidth]{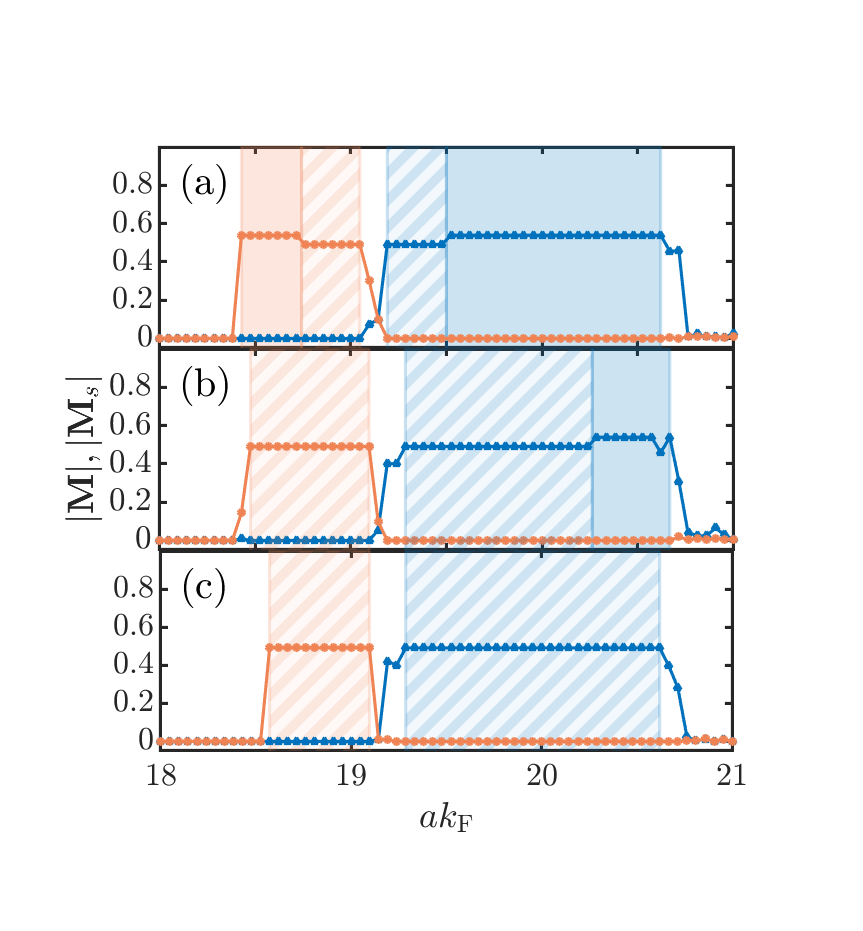}
    \caption{Phase diagram for the Ising spin symmetry with both dipole and RKKY interactions present, FM (AFM) denoted by blue (orange), and cluster-wise order 111 ($\tau10$) by striped (single-color) boxes. Parameters: (a) $J_\mathrm{D}/J_\mathrm{RKKY} = 0.1$, (b) $J_\mathrm{D}/J_\mathrm{RKKY} = 0.3$, (c) $J_\mathrm{D}/J_\mathrm{RKKY} = 0.5$.}
    \label{fig:ising_rkky_dipole_diagram}
\end{figure}

\begin{figure}[tb]
    \centering
    \includegraphics[width = 0.99\linewidth]{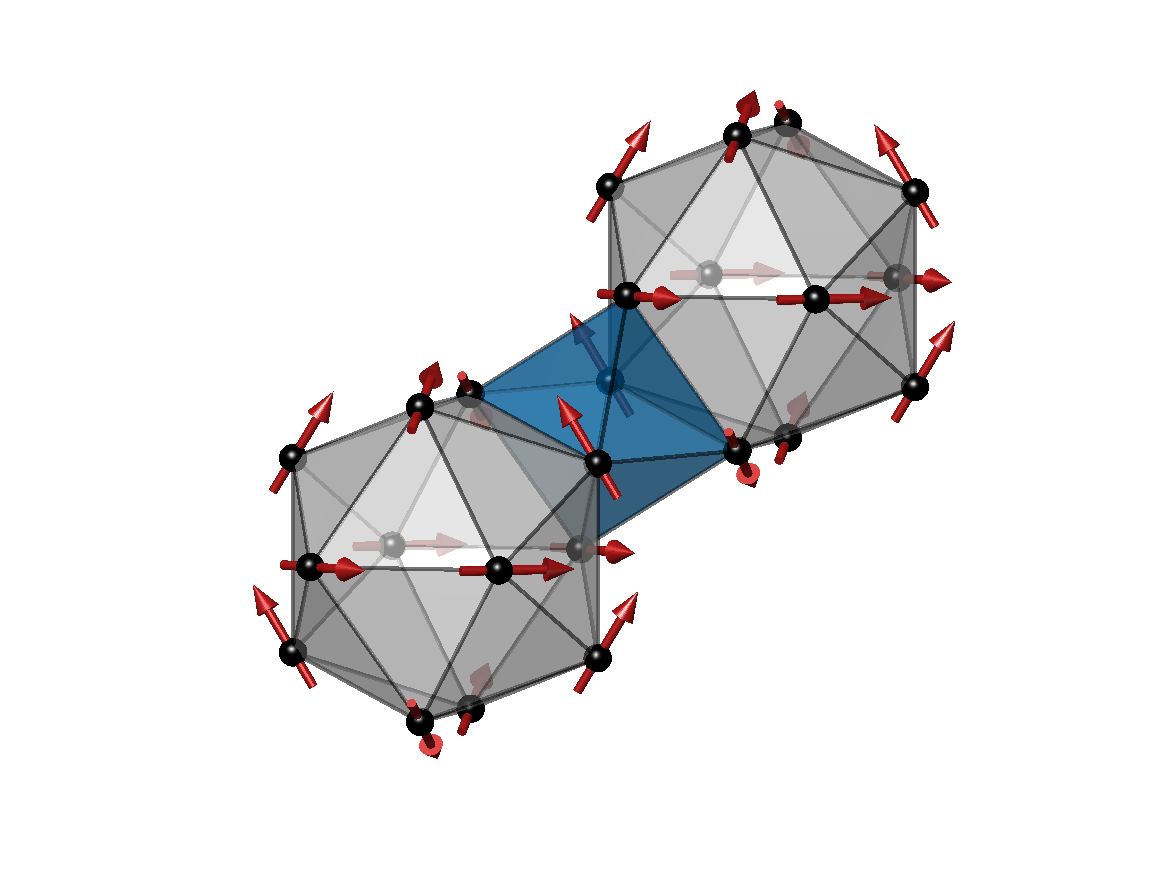}
    \caption{The FM 111 state, so denoted since the net magnetization direction is along the 111-direction. Total magnetization at 49\% of the corresponding Heisenberg ferromagnet. We bring attention to the ferromagnetic correlation between spins lying in the mirror planes of the lattice bisecting the icosahedra.}
    \label{fig:fm111}
\end{figure}

With a small dipole interaction alongside the RKKY interaction ($J_\text{D}/J_\text{RKKY} = 0.1$), the Ising model displays additional ground states ordered along the 111-directions, see Fig.~\ref{fig:ising_rkky_dipole_diagram} (a). The antiferromagnetic (AFM-111) and ferromagnetic (FM-111) states are related by flipping the spins located on the $B$-cluster, see Fig.~\ref{fig:geometry}. The FM-111 state is shown in Fig.~\ref{fig:fm111}, and resembles the ferromagnetic states found in Au-Si-Tb\cite{Hiroto_2020,Suzuki2021}. The AFM-111 state have been found in both Au-Si-Ho and Au-Si-Tb compounds \cite{PseudoTsai}, and in Au-Al-Tb\cite{SatoWhirling}. Increasing the dipole interaction to $J_\text{D}/J_\text{RKKY} = 0.5$, the states whose cluster-wise magnetization are aligned along the $\tau$10-directions almost disappear from the phase diagram, see Fig.~\ref{fig:ising_rkky_dipole_diagram} (c). We note here that a similar effect was found by Sugimoto et al.\cite{Sugimoto2016} by adding an on-site anisotropy term.

Considering combined RKKY and dipolar interactions under Heisenberg spin symmetry leads to a similar phase diagram as in Fig.~\ref{fig:mag_plot_rkky_dom}. However, the FM/AFM axes point along the 111-direction for arbitrarily small $J_\text{D}$, similar to Fig.~\ref{fig:dipole_gs_heisen}. Thus, for both Heisenberg and Ising spin symmetries, and with a dipole interaction stronger than roughly $J_\text{D}/J_\text{RKKY} = 0.5$, the ground states are eight-fold degenerate and based on our model we predict that approximants will have their magnetization axes along the 111-direction.

Experimentally, the transition between FM and AFM ground states occurs at larger electron concentrations, $\nu$, than those observed here. As seen in Fig.~\ref{fig:ising_rkky_dipole_diagram}, the critical $\nu_c$ is not affected by tuning the strength of the dipole interaction. Nonetheless, its presence seems to be important in order to acquire experimentally relevant ground states for this model.

\subsection{Interplay between RKKY and DE interactions}\label{subsec:kf_transition}
As neither anisotropy nor the dipole interaction was sufficient to explain the shift between Gd and Tb approximants, we turn to the interplay between the direct-exchange and RKKY terms. Here, focus lies on the only parameter in $\hamil_\text{DE}$ that modifies the observed magnetic order, $J_\text{O}$.

\begin{figure}[ht]
    \centering
    \includegraphics[width = 0.95\linewidth]{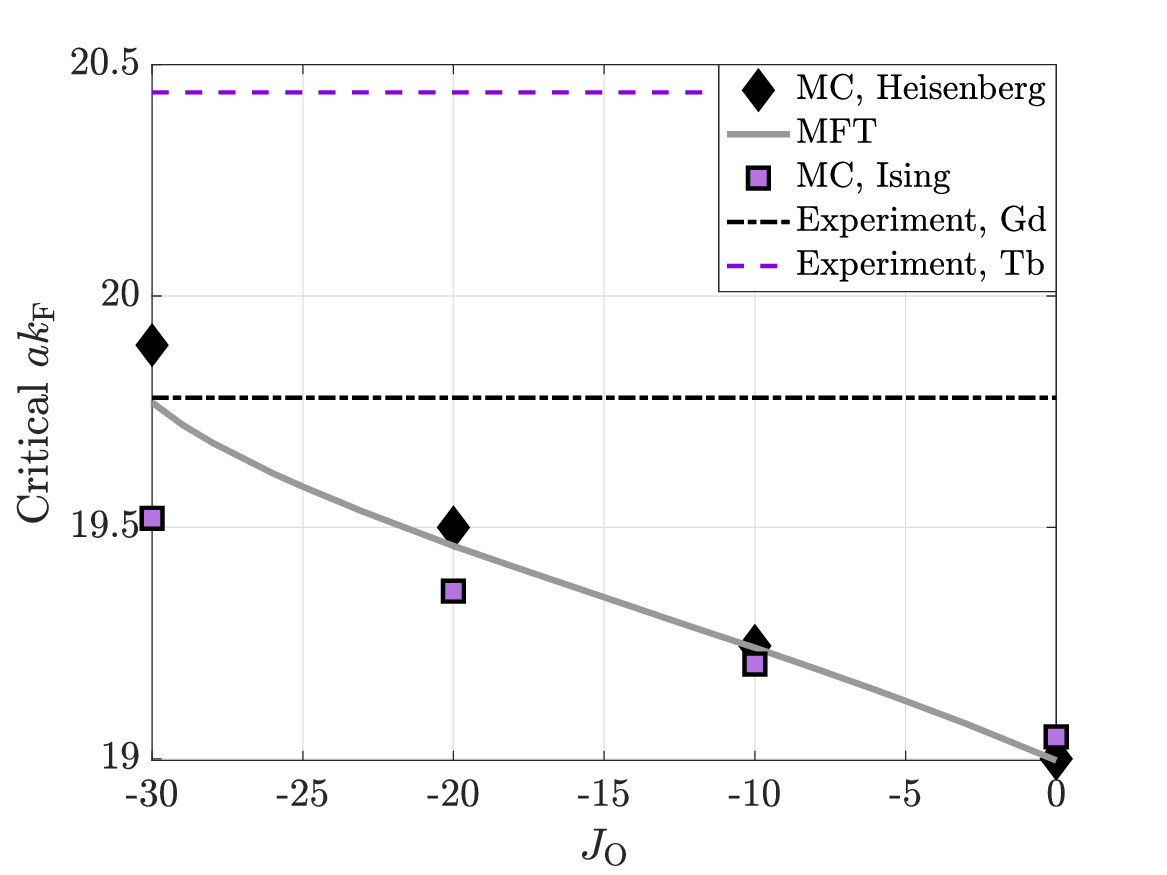}
    \caption{Critical values for $a k_\text{F}$ as a function of $J_\text{O}$. Mean field estimates from the point where $u_\text{AB}$ changes sign, while keeping $T_\text{c} > 0$, Eq.~(\ref{eq:tc_mft}). Since the experimental value for the critical $a k_\text{F}$ is beyond the reaches of the models studied here, we suggest that the band structure must be taken into account when computing the RKKY couplings.}
    \label{fig:akf}
\end{figure}

When the inter-sublattice direct-exchange term $J_\text{O} < 0$, it widens the antiferromagnetic region in Fig.~\ref{fig:mag_plot_rkky_dom}, and increases the critical value for the electron density. It also narrows the ferromagnetic region, in accordance with Eq.~(\ref{eq:tc_mft}). However, assuming the relationship between $k_\text{F}$ and $\nu$ in Eq.~(\ref{eq:kf_to_nu}), the transition point is still for smaller $\nu$ than experimentally observed. As seen in Fig.~\ref{fig:akf}, both the Heisenberg and Ising spin symmetries studied here follow roughly the estimate of the transition point computed from the mean-field theory presented in Eq.~(\ref{eq:tc_mft}). The drifts for larger amplitudes $|J_\text{O}|$ are expected: the behavior should be more mean-field like since the effective coordination number is large when $H_\text{RKKY}$ dominates, as long as the model does not become frustrated. The discrepancy in the Ising case can be expected given the assumption of isotropy in deriving the mean-field results, and the fact that fixed relative angles between nearest-neighbor easy axes reduce the effective impact of $J_\text{O}$ by a factor of $\cos \theta = \tau/(1 + \tau^2) \approx 0.45$. 

The different scaling of the critical Fermi wavevector versus $J_\text{O}$ for the Heisenberg and Ising spin symmetries could explain the shift in the AFM/FM transition point between Gd- and Tb-based materials if $J_\text{O} > 0$ and taken as equal for the different materials. However, this further pushes the transition point away from the experimental value. Another explanation in the framework of our model could be that Gd- and Tb-based materials have a different $J_\text{O}$. 

For $J_\text{O} \lesssim -10$, the FM (AFM) ground states for the Ising case have their magnetization (staggered magnetization) ordered along the $\tau 1 0 $-directions, and instead align along the $111$-directions. The total magnetization per spin is again found to be $49\%$ of that of a free ion. Thus, the addition of a nearest-neighbor interaction mimics the effect of including a dipole interaction and yields ground states reminiscent of the experimental spin structures observed in Tb-based approximant materials. 

\section{Conclusions}

We present a systematic study of the classical magnetic properties of the 1/1 Tsai-type approximant. The key observation is that the spin symmetry in the form of Ising or Heisenberg spins does not significantly affect the Fermi wavevector at which the ground state switches from an antiferromagnet to a ferromagnet when considering the RKKY interaction. This suggests that the differences between Gd-based and Tb-based approximants stem from another source. Our results also show that the difference is most likely not due to the dipolar intraction, and a remaining possible explanation within our model is a material dependent short-range interaction. In addition, the free electron model might be insufficient to model the conduction band.

For the approximant we find that the key parameter determining whether the ground state is ferro- or antiferromagnetic is not the much used Curie-Weiss temperature, but rather the sign of the inter-sublattice mean-field $u_\text{AB}$. This is the case for the direct exchange interaction, for which $u_\text{AB}$ is proportional to the direct exchange parameter $J_\text{O}$, and for the RKKY interaction, where the $k_\text{F}$-dependence of $u_\text{AB}$ correlates with the ground-state magnetization of both Heisenberg and Ising spin symmetries. 

We also note that the ground state of the approximant under Heisenberg spin symmetry and pure dipolar interaction, here properly treated with Ewald boundary conditions, is a ferromagnet, with an associated plateau in the the susceptibility below the transition, similar to the experimental observation in the recent Gd-based quasicrystal observed by Tamura et al \cite{Tamura2021}. However, for Ising spin symmetry, the most significant effect of perturbing the RKKY Hamiltonian with a dipolar contribution is a slight change in the ground state configuration, and the transition point is remarkably stable with respect to dipolar perturbations.

In conclusion, our study provides evidence against the hypothesis that the crystal electric field along with the nearly-free electron description of the conduction band are sufficient to explain the ground state differences between terbium and gadolinium. It would therefore be interesting to investigate whether the anisotropy makes a difference when using a non-spherical Fermi surface to compute the RKKY couplings.

\appendix

\acknowledgements
We thank Cesar Pay Gomez and Johan Hellsvik for insightful discussions, and Simon "Raba" Ranefj\"ard and Miguel Francisco Martinez Miquel for valuable input on figure design. This work has been financially supported by the Knut and Alice Wallenberg foundation (grant number KAW 2018.0019).The simulations were performed on resources provided by the Swedish National Infrastructure for Computing (SNIC) at the Center for High Performance Computing (PDC) at the Royal Institute of Technology (KTH).

\bibliography{apssamp}% Produces the bibliography via BibTeX.

\end{document}